\documentstyle[12pt, fleqn]{article}
\def\double{\baselineskip 24pt \lineskip 14pt}

\def\be{\begin{equation}}
\def\ee{\end{equation}}
\def\bea{\begin{eqnarray}}
\def\eea{\end{eqnarray}}
\def\l{\label}
\def\re{\cite}
\def\re#1{{[\ref{#1}]}}

\def\theequation{\ksection.\arabic{equation}}

\def\ta{\tilde{a}}

\def\p{\partial}

\def\m{{\cal{M}}}

\def\ta{\tilde a}
\def\bp{\beta_+}
\def\bm{\beta_-}

\begin{document}
\begin{titlepage}

\vspace*{-62pt}
\begin{flushright}
{\small
FERMILAB--Pub--93/100-A\\
August 1993}
\end{flushright}

\vspace{1in}

\begin{center}
\Large
{\bf Solutions to the Wheeler-DeWitt Equation Inspired by  the String Effective
Action}

\vspace{.3in}

\normalsize

\large{James E. Lidsey$^1$}

\normalsize
\vspace{.6cm}

{\em NASA/Fermilab Astrophysics Center, \\
Fermi National Accelerator Laboratory, Batavia, IL~~60510-0500, U.S.A.}

\end{center}

\vspace{3cm}

\baselineskip=24pt
\begin{abstract}
\noindent

The Wheeler-DeWitt equation is derived from the bosonic sector of the
heterotic string effective action assuming a toroidal
compactification. The spatially closed, higher dimensional
Friedmann-Robertson-Walker (FRW) cosmology is investigated and a
suitable change of variables rewrites the equation in a canonical
form. Real- and imaginary-phase exact solutions are found and a method
of successive approximations is employed to find more general power
series solutions. The quantum cosmology of the Bianchi IX universe is
also investigated and a class of exact solutions is found.

\end{abstract}

\vspace{2cm}
\small{$^{1}$e-mail: jim@fnas09.fnal.gov}

\vspace{1.5cm} Submitted to {\em Class. Quantum Grav.}

\end{titlepage}

\double

\section{Introduction}
\setcounter{equation}{0}
\vspace{.3cm}

Superstring theory remains a promising candidate for a complete theory
of quantum gravity \re{GSW}.  If string theory is indeed correct, it
should describe the complete quantum history of the universe. However
a consistent second quantization of the superstring is not currently
available and to make progress one must consider the low-energy limit
of the effective field theory.  One approach adopted by a number of
authors has been to investigate the quantum cosmology of the
superstring by solving the corresponding Wheeler-DeWitt (WDW) equation
for the effective action [2,3].

String theory and the WDW equation may be related at a fundamental
level. It has been shown that the effective action, $S$, may be
written as an integral over a spatial hypersurface that does not
depend on cosmic time \re{GS1988}. This is significant because the WDW
equation may be viewed as a time-independent Schr\"odinger equation
and its solution in the WKB approximation is determined by the phase
factor $\exp (iS)$. Consistency therefore suggests that the action
should not depend on cosmic time and the above approach appears well
motivated.

Recently the Scherk-Schwarz \re{Sch1979} dimensional reduction
technique has been employed to construct a string effective action in
which $d$ internal dimensions are compactified onto a torus
\re{KM1993}. In this paper we search for solutions to the WDW equation
derived from this action. Broadly speaking a singular, oscillating
solution may be interpreted at the semiclassical level in terms of
Lorentzian four-geometries, whereas an exponential solution can
represent a quantum wormhole if appropriate boundary conditions are
satisfied. Hawking and Page \re{HP1990} have proposed that the
wavefunction $\Psi$ should be exponentially damped at infinity and
obey a suitable regularity condition when the spatial metric
degenerates (i.e. its radius $a$ vanishes) \re{Zhuk}. Presently the
precise form of such a condition is not clear, but it appears that the
wavefunction should vary as $\Psi \propto a^p$, where the exponent $p$
is a constant. The sign of $p$ is associated with the choice of factor
ordering and may be negative \re{AED1992}.

The derivation of the string effective action is summarized in Sect. 2
and the corresponding WDW equation for the spatially closed,
$(D+1)$-dimensional Friedmann-Robertson-Walker (FRW) space-time is
derived. Singular and non-singular solutions to this equation are
found in Sect. 3. The dimensionally reduced action is invariant under
a global $O(d,d)$ transformation and in principle this symmetry may be
employed to generate new solutions. Such a procedure is illustrated
within the context of the duality symmetry of string theory
\re{dua}. The WDW equation is rewritten in a simplified canonical form
after a suitable redefinition of the independent variables and this
allows a method of successive approximations to be developed in
Sect. 4.  Additional power series solutions are found. Some of these
series can be written in closed form and are exponentially damped at
infinity. The more general Bianchi IX (mixmaster) cosmology is
investigated in Sect. 5 and an exact solution is found that is sharply
peaked around the FRW universe at large three-geometries.

Units are chosen such that $\hbar=c=16\pi G=1$.

\vspace{.3cm}
\section{The Effective Action and the Wheeler-DeWitt Equation}
\setcounter{equation}{0}
\vspace{.3cm}

We first summarize the derivation of the string effective action
presented in \re{KM1993}.  At the tree level, the Euclidean action of
the gauge singlet, bosonic sector of the $\hat{D}$-dimensional
heterotic string is\footnote{A hat $\hat{}$ denotes quantities in the
$\hat{D}$-dimensional spacetime ${\cal{M}}$, where $\hat{D}=1+D+d$.}
\be
\l{hataction}
S_{\hat{g}}=  \int_{{\cal{M}}} d^{\hat{D}}x \sqrt{\hat{g}}e^{-\hat{\phi}}
\left[  - \hat{R}(\hat{g}) -(\hat{\nabla} \hat{\phi} )^2 +\frac{1}{12}
\hat{H}_{\hat{\mu}\hat{\nu}\hat{\rho}}\hat{H}^{\hat{\mu}\hat{\nu}\hat{\rho}}
\right],
\ee
where $\hat{\phi}$ is the dilaton field and $\hat{H}$ is the totally
antisymmetric 3-index field \re{FT1985}. In the technique of
dimensional reduction the universe is viewed as the product space
$\m={\cal{J}}\times {\cal{K}}$, where the $(D+1)$-dimensional
space-time ${\cal{J}}(x^{\rho})$ has metric $g_{\mu\nu}(x^{\rho})$ and
the $d$-dimensional internal space ${\cal{K}}(y^{\alpha})$ must be
Ricci flat if the matter fields are independent of its coordinates
$y^{\alpha}$. This is the case for the Calabi-Yau spaces often
considered in string theory, but for the purposes of the present work
it is sufficient to assume ${\cal{K}}$ is a torus,
i.e. ${\cal{K}}=S^1\times S^1 \times \ldots \times S^1$. The complete
metric is then
\be
\l{metric}
\hat{g}_{\hat{\mu}\hat{\nu}}  =     \left( \begin{array}{cc}  g_{\mu\nu}+
A_{\mu}^{(1)\gamma} A^{(1)}_{\nu\gamma} & A^{(1)}_{\mu\beta} \\
A^{(1)}_{\nu\alpha} & G_{\alpha\beta}
\end{array} \right),
\ee
where $G_{\alpha\beta}$ is the metric on ${\cal{K}}$. The effective action in
$(D+1)$-dimensions is
\bea
\l{effact}
{S}=\int d^{D+1} x \sqrt{g} e^{-\phi} \left[ -R -(\nabla \phi)^2
+\frac{1}{12}H_{\mu\nu\rho}H^{\mu\nu\rho} - \frac{1}{8} {\rm Tr} \left(
\nabla_{\mu} M^{-1} \nabla^{\mu} M \right) \right. \nonumber \\
\left.  + \frac{1}{4} {\cal{F}}^i_{\mu\nu} \left( M^{-1} \right)_{ij}
{\cal{F}}^{\mu\nu j} \right],
\eea
where
\be
A^{(2)}_{\mu\alpha} = \hat{B}_{\mu\alpha} + B_{\alpha\beta}A^{(1)\beta}_{\mu}
\ee
\be
\phi = \hat{\phi} -\frac{1}{2} \ln  {\rm det} G
\ee
\be
H_{\mu\nu\rho}=\nabla_{\mu} B_{\nu\rho} - \frac{1}{2} {\cal{A}}_{\mu}^i
\eta_{ij} {\cal{F}}^j_{\nu\rho}+({\rm cyc.  perms.})
\ee
\be
{\cal{F}}^i_{\mu\nu} =  \nabla_{\mu} {\cal{A}}^i_{\nu} - \nabla_{\nu}
{\cal{A}}^i_{\mu}
\ee
and the $2d\times 2d$ matrices $M$,  $\eta$ and $M^{-1}$ are defined by
\be
\l{matrices}
M=\left( \begin{array}{cc} G^{-1} & -G^{-1}B \\ BG^{-1} & G-BG^{-1}B
\end{array} \right), \qquad \eta = \left( \begin{array}{cc} 0 & 1 \\ 1 & 0
\end{array} \right), \qquad M^{-1}=\eta M\eta ,
\ee
respectively. It follows from the definition of $M$ that $M^{\rm
T}\eta M = \eta$ and this implies $M \in O(d,d)$. Hence, since
$g_{\mu\nu}$ and $\phi$ are also invariant under the action of this
group, the dimensionally reduced action is symmetric under a global
$O(d,d)$ transformation.

We follow the approach of \re{KM1993} by choosing $\phi = {\rm
constant}$ with $H_{\mu\nu\rho}=0$ and ${\cal{A}}^i_{\mu} =0$.  We
assume a line element for ${\cal{J}}$ of the form
\be
ds^2 = n^2(t)dt^2 + a^2(t)d\Omega_D^2,
\ee
where $n(t)$ defines the lapse function, $d\Omega_D^2$ is the line
element of the unit $D$-sphere and $a(t)$ is the scale factor. This
assumption of spherical symmetry implies that all variables are
functions of $t$ only. Thus the $d\times d$ matrix $G+B$ may be
written as
\be
\l{G+B}
G+B = {\rm diag} \left( \Sigma_1, \ldots, \Sigma_{d/2} \right), \qquad \Sigma_j
= \left( \begin{array}{cc} e^{\psi_j} & \sigma_j \\ -\sigma_j & e^{\psi_j}
\end{array} \right) ,
\ee
where $G$ and $B$ are space-time dependent and we assume $d$ is
even. The action (\ref{effact}) then simplifies to
\be
\l{action}
S= \int d^{D+1} x \sqrt{g} \left( -R +\frac{1}{2} \sum_{j=1}^{d/2} \left[
(\nabla \psi_j )^2 + e^{-2\psi_j} ( \nabla \sigma_j )^2 \right] \right) .
\ee

 One special example of an $O(d,d)$ invariance is the {\em duality}
symmetry associated with an interchange of $M$ and $M^{-1}$
\re{dua}. The duality transformed fields are
\bea
\label{dualfields}
e^{-\tilde{\psi}_j} =   e^{\psi_j} + e^{-\psi_j}\sigma^2_j  \nonumber \\
\tilde{\sigma_j} = - \left( e^{\psi_j} + e^{-\psi_j}\sigma_j^2 \right)^{-1}
e^{- \psi_j} \sigma_j
\eea
and a new class of solution parametrized by $\{\tilde{\psi}_j,
\tilde{\sigma}_j \}$ may be generated from a solution given in terms
of $\{\psi_j, \sigma_j\}$.

The WDW equation for the theory (\ref{action}) may now be derived
using the standard techniques \re{KT1990}. The classical Hamiltonian
constraint is
\be
\l{classham}
a^D {\cal{H}} \propto  -a^2 \Pi_a^2 + 2D(D-1) \sum_{j=1}^{d/2} \left[
\Pi^2_{\psi_j} + e^{2\psi_j}\Pi^2_{\sigma_j} \right] - \alpha^2a^{2(D-1)} =0,
\ee
where ${\cal{H}}$ is the Hamiltonian density, $\Pi_{z_i} \equiv
\partial {\cal{L}}/\partial \dot{z}_i$ are the momenta conjugate to
$z_i$ and a dot denotes differentiation with respect to $t$. The
constant $\alpha = 2D(D-1)\Omega_D$ where $\Omega_D$ is the volume of
the unit $D$-sphere. The system is quantized by identifying the
conjugate momenta with the operators
\be
\l{quan}
\Pi_{a}^2 = -   a^{-p}  \frac{\partial}{\partial a}  {a}^p
\frac{\partial}{\partial a}, \qquad
\Pi_{\psi_j}^2 = - \psi_j^{-q_j} \frac{\partial}{\partial \psi_j} \psi_j^{q_j}
\frac{\partial}{\partial \psi_j}, \qquad
\Pi^2_{\sigma_j} =- \sigma_j^{-r_j}\frac{\partial}{\partial \sigma_j}
\sigma_j^{r_j} \frac{\partial}{\partial \sigma_j},
\ee
where the arbitrary constants $\{p, q_j, r_j \}$ account for
ambiguities in the operator ordering. The WDW equation follows by
viewing ${\cal{H}}$ as an operator acting on the wavefunction $\Psi [
a, \psi_j, \sigma_j ]$:
\bea
\l{WDW}
\left[  a^{2-p} \frac{\partial}{\partial a}a^p \frac{\partial}{\partial a}
- 2D(D-1) \sum_{j=1}^{d/2} \left( \psi_j^{-q_j} \frac{\partial}{\partial
\psi_j} \psi_j^{q_j} \frac{\partial}{\partial \psi_j} + e^{2\psi_j}
\sigma_j^{-r_j} \frac{\p}{\p \sigma_j} \sigma_j^{r_j} \frac{\p}{\p \sigma_j}
\right) \right. \nonumber \\
\left. - \alpha^2 a^{2(D-1)} \right] \Psi =0.
\eea
We proceed in the following sections to solve this equation.

\vspace{.3cm}
\section{Exact Solutions}
\setcounter{equation}{0}

\vspace{.3cm}

Exact solutions to Eq. (\ref{WDW}) may be found for $q_j=0$. We first
search for separable solutions of the form
\be
\l{sep}
\Psi[a,\psi_j,\sigma_j ] = \Phi [a,\Psi_j ] \prod_{j=1}^{d/2} C_j(\sigma_j).
\ee
The WDW equation decouples to  $(d+2)/2$ differential equations:
\be
\l{apsieqn}
\left[ a^{2-p} \frac{\p}{\p a} a^p \frac{\p}{\p a} - \alpha^2a^{2(D-1)} -
2D(D-1) \sum_{j=1}^{d/2} \left( \frac{\p^2}{\p \psi_j^2}
-\omega_j^2e^{2\psi_j}   \right) \right]\Phi =0
\ee
\be
\l{sigmaeqn}
\sigma_l^{-r_l} \frac{d}{d \sigma_l} \sigma_l^{r_l} \frac{dC_l}{d \sigma_l} +
\omega^2_lC_l =0, \qquad l=1, \ldots , d/2,
\ee
where $\omega_l$ are arbitrary separation constants. Eq. (\ref{sigmaeqn})
admits the exact solutions
\be
\l{sigmasol}
C_l = \sigma_l^{(1-r_l)/2}   J_{\frac{1-r_l}{2}} ( \omega_l \sigma_l), \qquad
\omega_l \ne 0
\ee
\be
\l{omega=0}
C_l = {\rm constant}, \qquad \omega_l=0,
\ee
where $J_p$ is a Bessel function of order $p$. Classically
Eq. (\ref{omega=0}) corresponds to $\dot{\sigma}_l=0$, which is a
particular solution to the Einstein field equations. In this case each
of the $\sigma_l$ may be set to zero without loss of generality by
means of a linear translation. Thus we shall first solve
Eq. (\ref{apsieqn}) for $\omega_l=0$. In this paper we are concerned
only with the functional forms of the solutions to Eq. (\ref{WDW}) and
we therefore ignore all constants of proportionality in the
wavefunction.

\vspace{.2cm}
\subsection{Case A: $\omega_l =0$}
\vspace{.2cm}

It proves convenient to make a change of variables to
 $\{u,v,s_1,\ldots , s_{(d-2)/2}\}$ {\em defined} in such a way that
 $\Phi$ is independent of the $s_j$. This necessarily results in some
 loss of generality but allows analytical solutions to be found. It is
 shown in the appendix that with a factor ordering $p=1$,
 Eq. (\ref{apsieqn}) reduces to the simplified canonical form
\be
\l{canonical}
\left[ \frac{\p}{\p u} \frac{\p}{\p v} - \frac{\alpha^2}{4} (uv)^{D-2} \right]
\Phi =0,
\ee
where
\be
\l{u}
u \equiv a \exp \left[ - \frac{1}{\sqrt{2D(D-1)}} \sum_{j=1}^{d/2} \psi_j \cos
\theta_j \right]
\ee
\be
\l{v}
 v \equiv a \exp \left[ \frac{1}{\sqrt{2D(D-1)}} \sum_{j=1}^{d/2} \psi_j \cos
\theta_j \right]
\ee
and the set of constants $\{\theta_j\}$ are solutions to the constraint
equation
\be
\l{constraint}
\sum_{j=1}^{d/2} \cos^2 \theta_j =1.
\ee
The $\{u,v\}$ variables are useful because they eliminate the direct
dependence of the WDW equation on the number of axion fields
present. We shall refer to them as the {\em canonical}
coordinates. These variables are the null coordinates over
minisuperspace when only one axion field is present \re{HP1986}. The
form of Eq. (\ref{canonical}) remains invariant if one adds or removes
extra massless scalar fields and the alterations are accounted for by
changing the number of fields included in the summations of
Eqs. (\ref{u}) -- (\ref{constraint}). We note also that
Eq. (\ref{canonical}) is invariant under an interchange of these
variables.

The variables $u$ and $v$ may be redefined as
\be
\l{xy}
u=\beta(y-x)^{2/(D-1)}, \qquad v=\beta(y+x)^{2/(D-1)} , \qquad \beta = \left(
\frac{D-1}{2 \alpha} \right)^{1/(D-1)}.
\ee
In this case   Eq. (\ref{canonical})  transforms to
\be
\l{wdwho}
\left[ \frac{\p^2}{\p y^2} - \frac{\p^2}{\p x^2} - y^2 + x^2 \right] \Phi =0,
\ee
which  is the equation for two harmonic oscillators with equal and opposite
energy \re{HP1990} .

A singular solution  to Eq. (\ref{apsieqn}) is
\be
\l{sing}
\Phi_{\gamma} =  K_{\pm i \epsilon} \left( \alpha a^{D-1}/(D-1) \right) \exp
\left[  i \gamma \sum_{j=1}^{d/2}   \psi_j \cos \Theta_j \right] , \nonumber
\ee
where
\be
\l{Thetaconstraint}
 \sum_{j=1}^{d/2} \cos^2 \Theta_j =1  ,
\ee
 $K$ is a modified Bessel function, $\gamma$ is an arbitrary constant and
\be
\l{epsilon}
\frac{\gamma}{\epsilon} \equiv \sqrt{ \frac{D-1}{2D}}  .
\ee
On the other hand, non-singular solutions to Eqs. (\ref{canonical}) and
(\ref{wdwho}) are
\be
\l{cont}
\Phi_c = \exp \left[ - \frac{\alpha}{2(D-1)} \left( c^{-1} u^{D-1} +cv^{D-1}
\right) \right]
\ee
and
\be
\l{hosol}
\Phi_n = \frac{1}{2^n n!} H_n(x)H_n(y) \exp \left[ -\left( x^2+y^2 \right)/2
\right] ,
\ee
respectively, where $c$ is an arbitrary integration constant and $H_n$
are Hermite polynomials \re{HP1990}. Solution (\ref{cont}) may be
written directly in terms of the scale factor and axion fields:
\be
\l{cont1}
\Phi_{\lambda} = \exp \left\{ - \frac{\alpha}{D-1} a^{D-1} \cosh \left[
\frac{\gamma}{\epsilon} \left( \sum_{j=1}^{d/2} \psi_j \cos \theta_j \right) +
\lambda \right] \right\} ,
\ee
where $\lambda \equiv \ln c$. Solutions equivalent to (\ref{hosol})
and (\ref{cont1}) have been found previously in a different context by
Zhuk \re{Zhuk1992}. They do not oscillate as the spatial metric
degenerates and they are exponentially damped at
infinity. Consequently they may be viewed as quantum wormholes with
asymptotic topology $\Re \times S^D$.

The ground state wavefunction of Eq. (\ref{cont}) corresponds to $c=1$
$(\lambda =0)$ and excited states, corresponding to $c\ne 1$, are
generated by linear translations on the $\psi_j$ fields
\re{Zhuk}. Thus there exists a continuous spectrum of these excited
wormhole states. On the other hand Eq. (\ref{hosol}) represents a
descrete spectrum of excited states \re{HP1990}, but the two spectra
have the equivalent ground state.  The duality transformation
(\ref{dualfields}) maps $\psi_j$ to $-\psi_j$ when the $\sigma_j$
fields vanish and this is equivalent to an interchange of the
canonical coordinates. Hence it follows from Eq. (\ref{cont}) that the
ground state may be interpreted as the self-dual solution to the WDW
equation. In this sense the ground state is associated with the
solution of maximum symmetry. In figure 1 this ground state and an
excited state of the continuous wormhole spectrum (\ref{cont}) are
plotted to illustrate this feature.

\vspace{.35cm}
\centerline{\bf Figure 1}
\vspace{.35cm}

Additional wormhole solutions may be found by factoring out the
continuous spectrum given by Eq. (\ref{cont}). If we write
\be
\l{****}
\Phi(u,v) \equiv \Delta(u,v)  \Phi_c(u,v),
\ee
it follows after substitution into Eq. (\ref{canonical}) that
Eq. (\ref{****}) is a solution to the WDW equation if $\Delta$
satisfies
\be
\l{eqnexp}
\frac{\p^2 \Delta}{\p u \p v} -  \frac{\alpha cv^{D-2}}{2} \frac{\p \Delta}{\p
u} - \frac{\alpha u^{D-2}}{2c} \frac{\p \Delta}{\p v} =0.
\ee
When $c=1$ a comparison with Eq. (\ref{hosol}) implies that $\Delta$
may be written as a product of two Hermite polynomials. For $c \ne 1$
one solution to Eq. (\ref{eqnexp}) is
\be
\Delta =   cv^{D-1} - c^{-1} u^{D-1}
\ee
\be
\quad = 2   a^{D-1} \sinh \left[  \frac{\gamma}{\epsilon} \left(
\sum_{j=1}^{d/2}  \psi_j \cos \theta_j \right) + \lambda \right] .
\ee

\vspace{.2cm}
\subsection{Case B:  $\omega_l \ne 0$}
\vspace{.2cm}

The question now arises as to whether singular and non-singular
solutions to Eq. (\ref{apsieqn}) can be found when $\omega_l \ne
0$. If we assume the separable ansatz
\be
\l{form for phi}
\Phi[a,\psi_j] = A(a) \prod_{j=1}^{d/2} B_j(\psi_j)
\ee
for arbitrary functions $A(a)$ and $B_j(\psi_j)$, we find
\be
\l{eqn for A}
a^{2-p}\frac{d}{da}a^p\frac{dA}{da} + \left( 2D(D-1)z^2 - \alpha^2 a^{2(D-1)}
\right) A=0
\ee
and
\be
\l{eqn for B}
\frac{1}{B_j} \frac{d^2 B_j}{d\psi^2_j} -  \omega_j^2 e^{2\psi_j} =-z^2_j =
-z^2 \cos ^2 \Theta_j,
\ee
where $\Theta_j$ again satisfy the integrability condition
(\ref{Thetaconstraint}) and $z$ is an arbitrary constant.

Eq. (\ref{eqn for B}) reduces to a Bessel equation after the change of
variables $\psi_j = \ln \xi_j$. Hence one solution to Eq. (\ref{apsieqn}) is
\be
\l{Phisol}
\Phi_z   = a^{(1-p)/2} K_{\pm s} \left( \alpha a^{D-1}/(D-1) \right)
\prod_{j=1}^{d/2}   K_{\pm i z_j} \left( \omega_j e^{\psi_j} \right) ,
\ee
where
\be
\l{s}
s^2=\frac{1}{(D-1)^2}\left[ \left( \frac{1-p}{2} \right)^2 - 2D(D-1)  z^2
\right]
\ee
and the   expression for the full wavefunction $\Psi$ is
\be
\l{sol for Psi}
\Psi[a, \psi_j, \sigma_j ] = a^{(1-p)/2} K_{s} \left( \alpha  a^{D-1}/(D-1)
\right) \prod_{j=1}^{d/2} \left\{ K_{iz_j} \left( \omega_j e^{\psi_j}
\right) J_{\frac{1-r_j}{2}} \left( \omega_j \sigma_j \right)
\sigma_j^{(1-r_j)/2} \right\} .
\ee

For simplicity let us consider the case $p=1$ and $r_j=0$. This leaves
the WDW equation invariant under a change of minisuperspace
coordinates \re{HP1986} and the $C_j$ functions represent plane wave
solutions. The modified Bessel function $K_q(x)$ is exponentially
damped at infinity, but diverges as $ K_q \propto x^{-q}$ in the limit
$x \rightarrow 0$. It follows from Eq. (\ref{Phisol}) that
\be
\l{limit}
\Phi_z \propto \exp \left[ -iz \left( \sqrt{2D(D-1)} \ln a  + \sum_{j=1}^{d/2}
\psi_j \cos \Theta_j \right) \right]
\ee
as $\{a, e^{\psi_j} \} \rightarrow 0$ and the wavefunction oscillates
an infinite number of times. This corresponds to an initial
singularity, but may be removed by integrating over the separation
constant $z$. This is analogous to taking the Fourier transform
\re{Zhuk}. The Riemann-Lebesgue lemma \re{BO1984} states that
\be
\l{rll}
\lim_{p \rightarrow + \infty} \int^{+\infty}_{-\infty} dx e^{ipx}=0
\ee
and it follows that the transformed wavefunction
\be
\l{phitilde}
\tilde{\Phi} = \int^{+\infty}_{-\infty} dz \Phi_z
\ee
is  damped both at the origin and at infinity. It is therefore Euclidean for
all values of $a$.

Finally new solutions may now be generated from Eq. (\ref{sol for
Psi}) in terms of the duality transformed fields defined in
Eq. (\ref{dualfields}). After a simple rearrangement we find
\be
\l{*}
e^{\psi_j} = e^{-\tilde{\psi}_j} \left( 1+\tilde{\sigma}_j^2
e^{-2\tilde{\psi}_j} \right)^{-1}
\ee
and
\be
\l{**}
\sigma_j=-\tilde{\sigma}_je^{-2\tilde{\psi}_j} \left( 1+ \tilde{\sigma}_j^2
e^{-2\tilde{\psi}_j} \right)^{-1} .
\ee
Substitution of these expressions into Eq. (\ref{sol for Psi})
generates a more complicated class of singular solution $\Psi[a,
\tilde{\psi}_j, \tilde{\sigma}_j]$.

\vspace{.3cm}
\section{Power Series Solutions via a Method of Successive Approximations}
\setcounter{equation}{0}
\vspace{.3cm}

Additional solutions to Eq. (\ref{canonical}) may be derived by
employing a method of successive approximations. We assume the
wavefunction may be expanded as the infinite sum
\be
\l{approx}
\Phi(u,v)  \equiv \sum_{m=0}^{\infty} \Phi_m(u,v)  .
\ee
This is a solution to  Eq. (\ref{canonical}) if  the functions $\Phi_0$ and
$\Phi_m$ $(m\ge 1)$  satisfy
\be
\l{approx1}
\frac{\p^2 \Phi_0}{\p u\p v} = 0
\ee
and
\be
\l{approx2}
\frac{4}{\Phi_{m-1}} \frac{\p^2 \Phi_m}{\p u \p v} = \alpha^2  (uv)^{D-2},
\qquad m\ge 1 ,
\ee
respectively. Eq. (\ref{approx1}) is the canonical form of the one-dimensional
wave equation and admits the {\em general} solution
\be
\l{approx3}
\Phi_0 = f(u) + g(v),
\ee
where $f(u)$ and $g(v)$ are arbitrary twice continuously differentiable
functions.

An iteration procedure may now be established by substituting the
solution for $\Phi_0$ into Eq. (\ref{approx2}), solving for $\Phi_1$,
and then repeating the process. The symmetry of Eqs.  (\ref{approx2})
and (\ref{approx3}) allows this to be done analytically by expressing
the solutions in terms of quadratures with respect to the canonical
coordinates. For simplicity we assume for the moment that $g(v)=0$. It
follows that for $m\ge 1$ $\Phi_m$ may be written in terms of
quadratures with respect to $u$:
\be
\l{approx4}
\Phi_m = \left( \frac{\alpha}{2} \right)^{2m} \frac{v^{m(D-1)}}{m!(D-1)^m}
\int^u du_m u_m^{D-2} \int^{u_m} du_{m-1} u_{m-1}^{D-2} \ldots \int^{u_2} du_1
u_1^{D-2} f(u_1).
\ee
By symmetry however, if $\Phi_0=g(v)$, the form of $\Phi_m$ is
identical to Eq. (\ref{approx4}) but with $u$ replaced by $v$ and
$f(u_1)$ replaced by $g(v_1)$. A more general expression for the
wavefunction is therefore
\be
\l{approx5}
\Phi = \Phi_0 + \sum_{m=1}^{\infty}  \left( \frac{ \alpha}{2} \right)^{2m}
\frac{\left( uv \right)^{m(D-1)}}{m!(D-1)^m} \left[ F_m (u) \pm G_m(v) \right]
\ee
\be
\label{approx6}
F_m(u) \equiv \frac{1}{u^{m(D-1)}} \int^u du_m u_m^{D-2} \int^{u_m} du_{m-1}
u_{m-1}^{D-2} \ldots \int^{u_2} du_1 u_1^{D-2} f(u_1)
\ee
\be
\l{approx7}
G_m(v) \equiv \frac{1}{v^{m(D-1)}} \int^v dv_m v_m^{D-2} \int^{v_m} dv_{m-1}
v_{m-1}^{D-2} \ldots \int^{v_2} dv_1 v_1^{D-2} g(v_1) .
\ee

The integrals in Eqs. (\ref{approx6}) and (\ref{approx7}) may be evaluated
analytically when
\be
\l{approx8}
\Phi_0 = v^b \pm u^b, \qquad b={\rm constant}
\ee
and it is straightforward to show  that
\be
\l{approx9}
F_m +G_m = \frac{\Phi_0}{[D-1+b][2(D-1)+b]\ldots [m(D-1)+b ]} .
\ee
It follows that the wavefunction reduces to the series solution
\be
\l{approx10}
\Phi = \Phi_0 \left[ 1+ \sum_{m=1}^{\infty} \left( \frac{\alpha}{2}
\right)^{2m} \frac{ \left( uv \right)^{m(D-1)}}{m!(D-1)^m} \frac{1}{[D-1+b]
\ldots [m(D-1)+b]} \right] .
\ee

\begin{table}
\begin{center}
\begin{tabular}{c||c|c}
 Solution    & $b$    & Wavefunction                   \\
\hline
\hline
             &        &                          \\
   I         &  $0$   & $I_0\left( \frac{\alpha (uv)^{(D-1)/2}}{D-1}\right) $
    \\
             &        &                          \\
   II        &  $\frac{D-1}{2}$  &  $\left( u^{-b} \pm v^{-b} \right) \sinh
\left( \frac{\alpha (uv)^{(D-1)/2}}{D-1} \right) $ \\
             &        &                           \\
    III      &  $\frac{1-D}{2}$  &  $\left( v^b \pm u^b \right) \cosh \left(
\frac{\alpha (uv)^{(D-1)/2}}{D-1}\right) $       \\
\end{tabular}
\end{center}
\footnotesize{\hspace*{0.2in} Table 1: Closed solutions to the canonical form
of the WDW equation
when $\Phi_0 = v^b \pm u^b$. $I_0$ is the modified Bessel function of
the first kind of order zero. All solutions diverge at infinity, but
linear combinations may be found which are exponentially damped at
infinity. It is straightforward to verify by differentiation that
these are solutions to Eq. (\ref{canonical}).}

\end{table}

Closed expressions for $\Phi$ exist when $b=0$ and $b=\pm (D-1)/2$ and
these solutions are shown in Table 1. They are Euclidean for all
values of the scale factor in the sense that they do not
oscillate. However they diverge exponentially fast at infinity and
therefore do not represent quantum wormholes in the Hawking-Page
sense. Solution I is identified as the Hartle-Hawking ground state
\re{HH1983}. On the other hand a linear combination of solutions II
and III is
\bea
\l{approx11}
\Phi_{\rm III} - \Phi_{\rm II} = \left( v^{-(D-1)/2} + u^{-(D-1)/2} \right)
\exp \left( - \frac{\alpha (uv)^{(D-1)/2}}{D-1} \right) \nonumber \\
  = \left( \frac{8\alpha}{\pi (D-1)} \right)^{1/2} K_{1/2} \left(
\frac{\alpha}{D-1} a^{D-1} \right) \cosh \left[ \left( \frac{D-1}{8D}
\right)^{1/2} \sum_{j=1}^{d/2} \psi_j \cos \theta_j \right] .
\eea
This solution has the correct asymptotic behaviour at infinity to be
interpreted as a quantum wormhole. Although it diverges as
$a^{(1-D)/2}$ when the spatial metric degenerates, this may be
acceptable behaviour for a quantum wormhole \re{AED1992}. Moreover,
Eq. (\ref{canonical}) was derived from the field theoretic limit of
the string effective action, and one might expect the analysis of
Sect. 2 to be invalid for scales below the Planck length. One could
therefore argue that this divergence will not arise in a more accurate
analysis.

\vspace{.35cm}
\section{An Exact Solution in the  Bianchi IX Cosmology}
\setcounter{equation}{0}
\vspace{.35cm}

The dimensionally reduced action (\ref{action}) is essentially
$(D+1)$-dimensional Einstein gravity minimally coupled to a set of
massless scalar fields with non-standard kinetic terms. It is
therefore possible when $D=3$ to extend the results from the previous
sections to the class of anisotropic and homogeneous Bianchi
cosmologies. The line element for the class A Bianchi spaces is
\re{EM1969}
\be
\l{line}
ds^2=-dt^2+e^{2 \ta (t)}  \left( e^{2 \beta (t)} \right)_{ij} \epsilon^i
\epsilon^j ,
\ee
where $e^{6 \ta}$ is the determinant of the metric on the
three-surface, $\beta_{ij}$ is a tracefree, symmetric matrix and
$\epsilon^j$ are the one-forms determining the isometry of the
three-surface. Without loss of generality $\beta_{ij}$ may be
diagonalized:
\be
\beta_{ij}(t)={\rm diag} \left[  \bp (t)+\sqrt{3}\bm (t), \bp (t) -\sqrt{3} \bm
(t), -2\bp (t) \right] .
\ee

Following an identical procedure to that presented at the end of
Sect. 2, the WDW equation derived from the action (\ref{action}) is
found to be
\be
\l{aniswd}
\left[ e^{-p\ta } \frac{\p}{\p \ta }e^{p\ta } \frac{\p}{\p \ta}
-\frac{\p^2}{\partial \bp^2} - \frac{\p^2}{\p \bm^2} + U\left( \ta, \beta_{\pm}
\right) - 12 \sum_{j=1}^{d/2} \left( \frac{\p^2}{\p \psi_j^2} + e^{2\psi_j}
\frac{\p^2}{\p \sigma_j^2} \right) \right] \Psi =0 ,
\ee
where the potential $U\left( \ta, \beta_{\pm} \right)$ depends on the
spatial curvature in the model. For simplicity we choose a factor
ordering $q_j=r_j=0$ for the scalar field momentum operators and
rescale such that $\alpha =1$. The WDW equation (\ref{WDW}) for the
positively curved, isotropic FRW model is recovered by setting $\bp
=\bm =0$ in the potential and removing the $\p^2 \Psi / \p
\beta_{\pm}^2$ terms in Eq. (\ref{aniswd}). When $\bm =0$ the Bianchi
IX universe simplifies to the Taub cosmology \re{T1951}.

To proceed we assume the separable ansatz:
\be
\l{***}
\Psi \left[ \ta, \bp , \bm, \psi_j, \sigma_j \right]  = \Phi \left[ \ta , \bp ,
\bm \right] S\left[ \psi_j , \sigma_j\right] .
\ee
The WDW equation separates into two differential equations:
\be
\l{sep1}
\left[ \frac{\p^2}{\p \ta^2 } +p \frac{\p}{\p \ta }  - \frac{\p^2}{\p \bp^2}
-\frac{\p^2}{\p \bm^2} + U  -m^2 \right] \Phi =0
\ee
\be
\l{sep2}
\left[ 12 \sum_{j=1}^{d/2} \left( \frac{\p^2}{\p \psi_j^2} + e^{2\psi_j}
\frac{\p^2}{\p \sigma_j^2} \right) -m^2 \right] S=0,
\ee
where $m$ is an arbitrary separation constant.

The curvature potential for the type IX may be written in the form \re{MTW1973}
\be
\l{curvpot}
 3 U  = e^{4\ta } \left[  e^{-8\bp} - 4e^{-2\bp} \cosh  2\sqrt{3}\bm
+2e^{4\bp} \left( \cosh  4\sqrt{3}\bm  -1 \right) \right] .
\ee
 It is well known that Eq. (\ref{curvpot}) exhibits a triangular
symmetry in the $\left(\bp , \bm \right)$ plane, but recently it was
shown \re{G1991} that there exists an additional symmetry
\be
\l{susy}
\left( \frac{\p \chi}{\p \ta} \right)^2 - \left( \frac{\p \chi}{\p \bp }
\right)^2 -\left( \frac{\p \chi}{\p \bm} \right)^2 + U =0,
\ee
where
\be
\l{phi}
\chi = \frac{1}{6} e^{2\ta }  {\rm Tr} \left( e^{2\beta} \right) = \frac{1}{6}
e^{2\ta } \left[ e^{-4 \bp } + 2e^{2\bp } \cosh    2\sqrt{3} \bm   \right] .
\ee

This additional symmetry may be employed to find new exact solutions.
If we substitute the secondary ansatz
\be
\l{5.10}
\Phi = W(\ta ) e^{-\chi}
\ee
into  Eq. (\ref{sep1}), and employ the identities
\be
\frac{\p^2 \chi}{\p \ta ^2} - \frac{\p^2 \chi}{\p \bp^2} - \frac{\p^2 \chi}{\p
\bm^2} +12 \chi = \frac{\p \chi}{\p \ta} -2\chi =0,
\ee
it follows that Eq. (\ref{sep1})  simplifies to
\be
\l{sep3}
\frac{\p^2 W}{\p \ta^2} +12 \chi W  +p\frac{\p W}{\p \ta } -2p\chi W -4 \chi
\frac{\p W}{\p \ta} -m^2 W=0.
\ee
For specific choices of the factor ordering parameter $p$,
Eq. (\ref{sep3}) admits the exponential solution $W=\exp ( - c \ta )$,
where $c$ is a decay constant.  Eq. (\ref{sep3}) reduces to the
constraint equation
\be
\l{c}
c^2 - pc -m^2 + 2(p-2c -6) \chi =0,
\ee
which is solved for
\be
p=2(c+3), \qquad c=-3 \pm \sqrt{(3+m)(3-m)} .
\ee
Requiring the factor ordering to be real implies $|m| \le 3$. This
solution is a generalization of the vacuum solution corresponding to
$m=0$ found in \re{MR1991}. We find a new vacuum solution for
$c=p=-6$.  It is worth remarking that the $e^{-\chi}$ factor in
Eq. (\ref{5.10}) appears to be independent of the choice of factor
ordering.

Finally we may combine these results with the solution to
Eq. (\ref{sep2}). This equation is solved by separating it into
Eqs. (\ref{sigmaeqn}) and (\ref{eqn for B}) as in Sect. 3. Hence, one
solution to Eq. (\ref{aniswd}) for the Bianchi IX cosmology is
\be
\l{IXsol}
\psi = e^{-c\ta - \chi}  \prod_{j=1}^{d/2} \left\{  K_{ \frac{m}{\sqrt{12}}
\cos \theta_j  }  \left( \omega_j e^{\psi_j} \right) e^{ \pm i \omega_j
\sigma_j} \right\} ,
\ee
where   $\omega_j$ are  arbitrary
constants that may be imaginary and the  $\theta_j$ must satisfy
$\sum_{j=1}^{d/2} \cos \theta_j^2=1$.

A new class of solution may now be generated from Eq. (\ref{IXsol}) as
a consquence of the duality symmetry of the action (\ref{action}), as
summarized in Eqs. (\ref{*}) and (\ref{**}). However, the real-phase
part of the solution, $\Phi$, remains invariant after such a
transformation.

When $c=0$ it was noticed in \re{MR1991} that as $\ta$ increases the
wavefunction $\Phi$ becomes peaked in the $(\bp , \bm )$ plane around
the isotropic FRW solution $\bp = \bm =0$. Our solutions also exhibit
this feature for values of $c$ given by Eq. (\ref{c}). In principle, a
solution of the form $\Phi = e^{-c\ta -\chi}$ exists for any WDW
equation that contains a separable component of the form
(\ref{sep1}). For example such a separation is possible with any
matter sector that behaves as a stiff perfect fluid at the classical
level. Dynamically this is equivalent to a massless scalar field and
at the quantum-mechanical level the matter is an eigenstate of some
operator with eigenvalue $m$. The constant $2\pi^2m$ represents a
conserved scalar flux.

\vspace{.3cm}
\section{Discussion}
\setcounter{equation}{0}
\vspace{.3cm}

In this paper the WDW equation was derived from the dimensionally
reduced string effective action assuming a toroidal
compactification. The $(D+1)$-dimensional FRW and the four-dimensional
Bianchi IX (mixmaster) cosmologies were investigated. In the former
example singular solutions were found and these may be interpreted as
spatially closed Lorentzian universes. Euclidean solutions were also
found which may be viewed as quantum wormholes if appropriate boundary
conditions are satisfied. The correct quantum theory of gravity should
explain, either from symmetry or probabilistic considerations, why
compactification to a four-dimensional space-time is observed. For
this reason the dimensionality of the external space was left
unspecified in the analysis.

In the Bianchi type IX case exact solutions were obtained for specific
choices of factor ordering. The wavefunction is strongly peaked around
the FRW universe at large three-geometries. Recently a symmetry of the
form (\ref{susy}) has been found for the Bianchi type II metric
\re{OSB1993} and it appears that this symmetry exists for all class A
spaces. In principle solutions of the form Eq. (\ref{IXsol}) can
therefore be found for these spaces.

In both the FRW and Bianchi IX examples the $O(d,d)$ symmetry of the
string effective action may be employed to generate new classes of
solutions which may otherwise be difficult to derive using standard
techniques.  This was illustrated using the duality symmetry of the
action, which is a subgroup of $O(d,d)$. The action (\ref{action}) is
also symmetric under the group ${\rm SL}(2,{\rm R})$ and new solutions
may also be generated from this symmetry \re{KM1993}.

A method of successive approximations was employed for the FRW
cosmology to derive further solutions. This method should be useful
for finding perturbative solutions when $uv = a^2$ is small (but not
so small that the analysis of Sect. 2 breaks down). In this region one
may view the power series as a truncated Taylor series up to some
appropriate order in the scale factor.

Finally, in the regime where the curvature term of
Eq. (\ref{canonical}) becomes negligible, the Wheeler- DeWitt equation
(\ref{canonical}) reduces to the one-dimensional wave equation:
\be
\l{k=0}
\frac{\p^2 \Phi}{\p u\p v} =0
\ee
with general solution $\Phi=f(u)+g(v)$.  Eq. (\ref{k=0}) may also be
treated as the exact WDW equation for a spatially flat Friedmann
universe with nontrivial spatial topology such as the 3-torus
$T^3=S^1\times S^1 \times S^1$ \re{Grish}. This illustrates a further
advantage of employing the canonical coordinates. In this case one may
easily find forms of $f(u)$ and $g(v)$ which satisfy the Hawking-Page
boundary conditions and this suggests that a positive spatial
curvature is not a necessary condition for the existence of quantum
wormholes \re{CS1993}.

\vspace{1cm}
{\bf Acknowledgments} The author is supported by the Science and
Engineering Research Council (SERC) U K and is supported at Fermilab
by the DOE and NASA under Grant NAGW-2381.

\newpage


\centerline{{\bf References}}

\begin{enumerate}

\item  \label{GSW}  Green M B,  Schwarz J H and Witten E 1988 {\em Superstring
Theory} (Cambridge: Cambridge University Press)

\item \label{W}   DeWitt B S 1967 {\em Phys. Rev.} {\bf 160}  1113

\item[]  Wheeler J A 1968 {\em   Battelle Rencontres} (New York: Benjamin)

\item \label{P}   Enqvist K,  Mohanty S  and   Nanopoulos D V 1987 {\em Phys.
Lett.} {\bf 192B}   327

\item[]  Enqvist K,   Mohanty S and Nanopoulos D V 1989 {\em Int. J. Mod.
Phys.} A {\bf  4} 873

\item[] Pollock M D 1989 {\em Nucl. Phys.} B {\bf 315}  528

\item[] Lyons A and   Hawking S W 1991 {\em Phys. Rev.}  D {\bf 44}   3802

\item[]  Pollock M D 1992 {\em Int. J. Mod. Phys.} A {\bf 7}   4149

\item[]   Wang J 1992 {\em Phys. Rev.} D {\bf 45} 412

\item \label{GS1988} Giddings S B and Strominger A 1988 {\em Nucl. Phys.} B
{\bf 306} 890

\item[] Pollock M D 1991 {\em Nucl. Phys.} B {\bf 355} 133

\item \label{Sch1979} Scherk J and Schwarz J H 1979 {\em Nucl. Phys.} B {\bf
153} 61

\item \label{KM1993} Khastgir S P and   Maharana J 1993 {\em Phys. Lett.} {\bf
301B}  191

\item[] Maharana J and Schwarz J H 1993 {\em Nucl. Phys.}  B {\bf 390} 3

\item \label{HP1990}   Hawking S W 1990  {\em Mod. Phys. Lett.} A {\bf 5}   453

\item[] Hawking S W and   Page D N 1990 {\em Phys. Rev.}  D {\bf 42}   2655

\item \label{Zhuk}    Garay L J 1991 {\em Phys. Rev.} D {\bf 44} 1059

\item[]   Kiefer C 1988 {\em Phys. Rev.} D {\bf 38} 1761

\item[]  Campbell L M and Garay L J 1991 {\em Phys. Lett.} {\bf 254B} 49

\item[]   Wada S 1992 {\em Mod. Phys. Lett.} A {\bf 7} 371

\item[]  Gonzalez-Diaz P F 1990 {\em Mod. Phys. Lett.} A {\bf 5} 1307

\item \label{AED1992} Alty L J, D'Eath P D and Dowker H F 1992 {\em Phys. Rev.}
D {\bf 46} 4402

\item \label{dua}   Kikkawa K and Yamasaki M 1984 {\em Phys. Lett.} {\bf 149B}
357

\item[]   Sakai N and   Senda I 1986 {\em Prog. Theor. Phys.} {\bf 75} 692

\item[]   Pollock M D 1989 {\em Phys. Lett.} {\bf 227B}   221

\item[]  Veneziano G 1991 {\em Phys. Lett.} {\bf  265B}   287

\item[]  Meissner  K A and   Veneziano G 1991 {\em Mod. Phys. Lett.} A {\bf 6}
 3397

\item \label{FT1985}   Fradkin E S  and   Tseytlin A A 1985 {\em Phys. Lett.}
{\bf 158B}  316; {\em Nucl.  Phys.} B {\bf 261}   1

\item[] Callan C G, Friedan D,  Martinec E J and   Perry M J 1985 {\em Nucl.
Phys.} B {\bf 262} 593

\item \label{KT1990} Kolb E W and Turner M S 1990 {\em The Early Universe} (New
York: Addison-Wesley)

\item \label{HP1986}   Hawking S W and   Page D N 1986 {\em Nucl. Phys.}  B
{\bf 264}  185

\item \label{Zhuk1992}  Zhuk A 1992 {\em Phys. Rev.} D {\bf 45} 1192

\item \label{BO1984} Bender C M and Orszag S A 1984 {\em Advanced Mathematical
Methods for Scientists and Engineers}, (Singapore: McGraw-Hill)

\item \label{HH1983}  Hartle J B and  Hawking S W 1983 {\em Phys. Rev.} D {\bf
28}  2960

\item[] \label{Zhuk2} Zhuk A 1992 {\em Class. Quantum Grav.} {\bf 9} 2029

\item \label{EM1969} Ellis G F R and MacCallum M A H 1969 {\em Commun. Math.
Phys.} {\bf 12} 108

\item \label{T1951} Taub A 1951 {\em Ann. Math. }{\bf 53} 472

\item \label{MTW1973} Misner C, Thorne K and Wheeler J A 1973 {\em
Gravitation}, (Freeman: San Francisco)

\item \label{G1991} Graham R 1991 {\em Phys. Rev. Lett.} {\bf 67} 1381

\item \label{MR1991} Moncrief V and Ryan Jr. M P 1991 {\em Phys. Rev.} D {\bf
44} 2375

\item \label{OSB1993} Obreg\'on O, Socorro J and Ben\'itez J 1993 {\em Phys.
Rev.} D {\bf 47} 4471

\item \label{Grish} Grishchuk L P and Sidorov Yu V 1987 {\em Proceedings on the
Fourth Seminar on Quantum Gravity}, (Singapore: World Scientific)

\item \label{CS1993} Coule D H and Solomons D 1992 University of Cape Town
Preprint

\end{enumerate}

\newpage

\appendix

\large{\centerline{\bf Appendix}}

\normalsize
\setcounter{equation}{0}

\def\theequation{A.\arabic{equation}}

\vspace{.3cm}

The purpose of this appendix is to derive the expressions for the canonical
coordinates when $\omega_j=0$. We  make an arbitrary change of variables in
Eq.  (\ref{apsieqn}) and search for solutions that depend on only two of these
new variables $u$ and $v$. We choose  $p=1$  and define $\ta = \ln a$ for
convenience.  The differential momentum operators of Eq. (\ref{apsieqn}) take
the form
\be
\frac{\p^2 \Phi}{\p w^2} = \frac{\p^2 u}{\p w^2}\frac{\p \Phi}{\p u} + \left(
\frac{\p u }{\p w} \right)^2 \frac{\p^2 \Phi}{\p u^2} + 2 \frac{\p u}{\p
w}\frac{\p v}{\p w} \frac{\p^2 \Phi}{\p u\p v} + \frac{\p^2 v}{\p w^2}\frac{\p
\Phi}{\p v} + \left( \frac{\p v}{\p w} \right)^2 \frac{\p^2 \Phi}{\p v^2},
\ee
where the variable $w$ is to be identified with $\ln a$ and $\psi_j$.

One may arrange for the majority of these terms to cancel after substitution
into Eq. (\ref{apsieqn}). Only the contributions containing $\p^2 \Phi/\p u\p
v$ remain if $u$ is a solution of the $d/2$-dimensional wave equation:
\be
\l{62}
\frac{\p^2 u}{\p \ta^2} - 2D(D-1) \sum_{j=1}^{d/2} \frac{\p^2 u }{\p \psi_j^2}
=0,
\ee
subject to the integrability condition
\be
\l{63}
\left( \frac{\p u}{\p \ta} \right)^2 - 2D(D-1) \sum_{j=1}^{d/2} \left( \frac{\p
u}{\p \psi_j} \right)^2 =0 .
\ee
Equivalent equations also apply  for $v$. The  first integral of Eq. (\ref{62})
is
\be
\frac{\partial u}{\partial \psi_j} = \pm \frac{\cos \theta_j }{\sqrt{2D(D-1)}}
\frac{\partial u}{\partial \tilde{a}} ,
\ee
where the constants of integration $\theta_j$ satisfy the constraint equation
$\sum_{j=1}^{d/2} \cos^2 \theta_j =1$. It follows that
\be
u=f\left( \ln a \pm \left[ 2D(D-1)\right]^{-1/2} \sum_{j=1}^{d/2} \psi_j \cos
\theta_j \right) ,
\ee
where $f$ is some arbitrary function. To ensure that the $\partial^2 \Psi /
\partial u \partial v$ terms do not cancel in the WDW equation, the argument in
 $v$ for the scalar fields must take the opposite sign to that in $u$, i.e.
\be
v=g\left( \ln a \mp \left[ 2D(D-1) \right]^{-1/2} \sum_{j=1}^{d/2} \psi_j \cos
\theta_j \right) ,
\ee
where $g$ is a second arbitrary function.

If we assume the separable ansatz
\be
u = u(\ta ) \prod_{j=1}^{d/2} u_j(\psi_j), \qquad v= v(\ta ) \prod_{j=1}^{d/2}
v_j(\psi_j ) ,
\ee
it is easy to verify by differentiation that one solution to these equations is
given by Eqs. (\ref{u}), (\ref{v}) and (\ref{constraint}).

\newpage

\centerline{\bf Figure Caption}

\vspace{2cm}

{\em Figure 1:}  In Figure (1a) a schematic plot of the self-dual ground state
of both the continuous and descrete wormhole spectra (\ref{cont}) and
(\ref{hosol}) is shown as a function of the canonical coordinates. The
symmetric nature of the solution is clearly illustrated. In Figure (1b) an
excited state of (\ref{cont})  is plotted for $c=1/2$.  As noted in the text
the ground state may be viewed as the solution of maximum symmetry.

\end{document}